\documentclass[prd,onecolumn,notitlepage,nofootinbib,superscriptaddress,showpacs,showkeys]{revtex4-2}

\usepackage{amsfonts}
\usepackage{graphicx}
\usepackage{dcolumn}
\usepackage{amsmath}
\usepackage{amssymb}
\usepackage[T1]{fontenc}
\usepackage[utf8]{inputenc}
\usepackage{esint}
\usepackage[brazil,english]{babel}
\usepackage[usenames,dvipsnames]{color}
\usepackage{longtable}
\usepackage{ulem}
\usepackage{nicefrac}
\usepackage{bm}
\usepackage{tikz}
\usepackage{braket}
\usepackage{graphicx, nicefrac}
\usepackage{cancel}
\usepackage{mathrsfs}
\usepackage{enumitem} 
\begin{document}
	
	\title{
	Quasi-local thermodynamics of Kerr--Newman black holes: Pressure, volume, and shear work
	}
	\author{T. L. Campos}
	\email{thiagocampos@alumni.usp.br}
\affiliation{Centro Brasileiro de Pesquisas Físicas, CEP 22290-180, Rio de Janeiro-RJ, Brazil}

	\begin{abstract}
		While the quasi-local thermodynamics of spherically symmetric black holes is well described by pressure and volume, extending this framework to rotating spacetimes poses a significant challenge. Rotation induces an oblate deformation of the horizon, breaking the direct functional dependence between geometric volume and area. In this work, we resolve this difficulty by establishing a quasi-local thermodynamic framework for Kerr-Newman black holes. We demonstrate that accommodating this kinematic deformation requires extending the thermodynamic phase space to include a geometric eccentricity parameter $Y$ and its conjugate, a thermodynamic shear tension $X$. Consequently, the rotational contribution is incorporated into the first law with a shear work term $X dY$. We derive the generalized first laws and Smarr formulas (Euler relations) for both the internal energy and enthalpy representations, showing that these thermodynamic potentials can be obtained through Legendre transformations that isolate the quasi-local energy from the rotational energy. Thus, this framework provides a novel perspective on the thermodynamics of rotating black holes, integrating the geometric deformation of the horizon into a quasi-local description.
	\end{abstract}
	
	\maketitle
	

\section{Introduction}
\label{sec:introduction}

A prominent avenue of research in the thermodynamic description of black holes is the extended phase space formalism, also known as black-hole chemistry \cite{mann2025black, kubizvnak2017black}, in which the cosmological constant is incorporated as a dynamical variable representing pressure, and its conjugate quantity is the thermodynamic volume. This framework has revealed profound connections between black-hole mechanics and classical thermodynamics \cite{kastor2009enthalpy, gibbons2005first, caldarelli2000thermodynamics, campos2024generating, campos2025black, xiao2024extended}, including van der Waals-like phase transitions and critical phenomena in anti-de Sitter spacetimes \cite{dolan2011compressibility, dolan2011pressure, dolan2012pdv, cvetivc2011black, kubizvnak2012p}.

Parallel to these developments, it is well established that general spherically symmetric black-hole horizons admit a quasi-local formulation of their thermodynamics \cite{hayward1994general, hayward1996gravitational, hayward1998unified, hayward2009local, padmanabhan2002classical, faraoni2015cosmological}. This formalism stands in contrast to traditional global characterizations, such as the ADM mass and the event horizon, which strictly depend on the asymptotic structure of the spacetime \cite{penrose1982some}. While these global features are convenient for formal definitions, they are also inherently teleological constructs, rendering them inaccessible to any realistic observer \cite{visser2014physical}.

In this quasi-local regime for spherically symmetric horizons, the first law assumes the traditional forms \mbox{$dU = T dS - P dV$} or \mbox{$dH = T dS + V dP$}, corresponding to the internal energy and enthalpy representations, respectively. Within this formulation, the thermodynamic potentials $U$ and $H$ generally do not coincide with the global asymptotic masses. 
On the other hand, extending this quasi-local structure to rotating horizons is nontrivial, as rotation deforms the horizon into an oblate geometry and breaks the simple relation between its area (entropy) and volume. Therefore, standard pressure-volume terms alone do not capture the full thermodynamic response. 

In this work, we address this problem for Kerr--Newman black holes by introducing a geometric eccentricity parameter~$Y$ and its conjugate thermodynamic shear tension $X$. The rotational deformation then contributes through a shear-work term, leading to the generalized first laws $dU = T dS - P dV + X dY$ and $dH = T dS + V dP + X dY$.

Our framework can be contextualized by comparison with the surface-tension approach to horizon thermodynamics proposed by Hansen, Kubiz\v{n}ák, and Mann~\cite{hansen2016criticality}, which has subsequently been applied to a variety of black-hole and cosmological horizons~\cite{chen2017modified, chen2019modifying, villalba2020kerr, jawad2018thermodynamic}. The surface-tension approach considers a first law of the form $d E = T dS + \Omega_H d J - \sigma dA,$ where $\sigma$ is a surface tension associated with the horizon area $A$. In that approach, $E$ is identified as a generalized Misner-Sharp mass by algebraically rearranging $T dS$, written in terms of $r_+$ and $a$, for a Boyer-Lindquist-type ansatz. Our framework differs in three important respects. First, our generalized Misner-Sharp mass~$U$ can be derived within the covariant phase space formalism. Second, rewriting the surface-tension contribution as a standard $-PdV$ work term while treating $r_+$ and $a$ as independent requires $V\propto A^{3/2}$, which does not coincide with the geometric volume of a rotating horizon, whereas our approach considers the volume within an ellipsoidal horizon, which can be rigorously obtained using Parikh's or vector-volume formalisms~\cite{parikh2006volume,ballik2013vector}. Finally, in the present paper, the mechanical work terms in the first law are directly associated with geometric deformations of the horizon, providing a natural interpretation of $U$ as internal energy. This, in turn, allows a transformation to the enthalpy representation, in which the thermodynamic variables are independent.

The paper is organized as follows. Section \ref{sec:quasi-local spherical} reframes the quasi-local thermodynamics of spherically symmetric horizons, laying the foundation for the subsequent developments. Section~\ref{sec:Iyer-Wald} generalizes the geometric formulation of internal energy to Kerr--Schild geometries within the covariant phase space formalism. Section \ref{sec:quasi-local Kerr--Newman} develops the extended quasi-local framework for Kerr--Newman, including the internal-energy and enthalpy representations, limiting cases, and the geometric interpretation of the deformation. Section \ref{sec:final_remarks} presents our concluding remarks. We adopt geometric units and signature $(-,+,+,+)$.


\section{Quasi-local thermodynamics of spherically symmetric horizons}
\label{sec:quasi-local spherical}

\subsection{The general formalism}

To set the stage for our axisymmetric generalization, we first recast the quasi-local thermodynamics of spherically symmetric horizons in a setting that accommodates the subsequent developments of this work.
The most general static spherically symmetric metric can be written in advanced Eddington--Finkelstein (EF) coordinates as
\begin{equation}
	ds^2 = -F(r) e^{2\psi(r)} dv^2 + 2e^{\psi(r)} dv dr + r^2 d\Omega^2~,
\end{equation}
where $v$ is the advanced time, $r$ is the areal radius, and $d\Omega^2$ is the standard metric of the unit two-sphere. The metric function is parameterized by the Misner-Sharp mass $M(r)$ as 
\begin{equation}
	F(r) = 1- \frac{2M(r)}{r}~.
\end{equation}
The black-hole horizon, formally a future outer trapping horizon, is located by the condition \mbox{$F(r_+) = 0$} with \mbox{$F'(r_+) > 0$}~\cite{hayward1994general, faraoni2015cosmological}.

From this geometric setup, a general first law at the horizon can be derived. The starting point is to evaluate
\begin{align}
	(dM)|_{F=0} &= \bigg[d\bigg( \frac{r}{2}\bigg) (1-F)-\frac{r}{2} dF\bigg]_{F=0} = d(M|_{F=0}) + \bigg[\frac{r}{2}d\bigg( \frac{2M}{r} \bigg)\bigg]_{F=0}~.
\end{align}
Using the condition $F(r_+)=0$ at the horizon radius $r_+$, we identify the surface gravity as
\begin{equation}
	-d\bigg( \frac{2M}{r} \bigg)\bigg|_{F=0} = 2 \kappa_+ dr|_{F=0}~,  \qquad \kappa_+ = \frac{1}{2} \frac{\partial F}{\partial r}(r_+)~.
\end{equation}
Hence, the differential of the Misner-Sharp mass at the horizon yields
\begin{equation}
	(dM)|_{F=0} = d(M|_{F=0}) - T dS~,
	\label{first law 0}
\end{equation}
with the standard geometric definitions for entropy and Hawking temperature:
\begin{equation}
	S = \pi r_+^2~, \qquad T = \frac{\kappa_+}{2\pi}~.
\end{equation}
Note that the temperature is strictly positive.

Alternatively, from the unified first law of Hayward's quasi-local mechanics \cite{hayward1994general, hayward1996gravitational, hayward1998unified, hayward2009local}, the dynamics of the trapping horizon follows
\begin{equation}
	(dM)|_{F=0} = \left( \partial_rM dr \right)_{F=0} = \left[ \left(A \psi_r + w \partial_r V\right)dr \right]_{F=0}~,
\end{equation}
where $\psi^\mu$ is the energy flux vector and $w$ is the work density scalar, constructed from the Einstein tensor $G_{\mu\nu}$ as
\begin{equation}
	\psi_\mu = \frac{1}{8\pi}G_\mu{}^{r} + w \delta_\mu^r~, \qquad w = - \frac{1}{16\pi} h^{ab} G_{ab}~,
\end{equation}
where $h_{ab}$ is the metric on the $(1+1)$-dimensional space orthogonal to the spherical orbits.
Through the Einstein field equations, these geometric quantities are directly linked to the matter content of the spacetime. In EF coordinates, the radial derivative of the Misner-Sharp mass is
\begin{equation}
	\partial_r M = - 4\pi r^2 (T_v{}^v-\rho_\Lambda)~,
	\label{partial r M}
\end{equation}
where $T_{\mu\nu}$ is the energy-momentum tensor and \mbox{$\rho_\Lambda = \Lambda/(8\pi)$} represents the energy density due to the cosmological constant~$\Lambda$. 

Within this framework, the thermodynamic pressure $P$ is defined for spherically symmetric geometries by evaluating the radial stress at the horizon,
\begin{equation}
	P \equiv \left(T_v{}^v-\rho_\Lambda\right)_{F=0} = \left(T^r{}_r-\rho_\Lambda\right)_{F=0}~.
	\label{PT}
\end{equation}
From Eq.~\eqref{partial r M},
\begin{equation}
	P = - \frac{\partial_r M}{4\pi r^2}\bigg|_{F=0}~.
\end{equation}

Therefore, equating the Hayward mechanical law with Eq.~\eqref{first law 0}, the first law of black-hole thermodynamics naturally emerges:
\begin{equation}
	dU = T dS - P dV~,
\end{equation}
where we identify the internal energy $U$ and the geometric volume $V$ of the black hole,
\begin{equation}
	U \equiv M|_{F=0} = \frac{r_+}{2}~, \qquad V = \frac{4\pi r_+^3}{3}~. \label{UMV}
\end{equation}
By recognizing $U(S,V)$ as a homogeneous function, we can apply Euler's theorem to obtain the generalized Smarr relation. The scaling dimensions are
\begin{equation}
	[U]\sim L~, \quad [S]\sim L^2~,\quad  [V]\sim L^3~,
	\label{scaling}
\end{equation} 
where $L$ denotes a length scale. It then follows that
\begin{equation}
	U = 2TS - 3PV~. \label{Euler scaling}
\end{equation}

We can also define an enthalpy representation via
\begin{equation}
	H = U + PV~,
\end{equation}
leading to
\begin{equation}
	dH = T dS + V dP~. \label{H_esf}
\end{equation}
Applying the Legendre transform to the Smarr relation, we find
\begin{equation}
	H = 2TS - 2PV~.
\end{equation}
In this representation, the enthalpy is a function of entropy and pressure, $H(S,P)$, taking the generalized form
\begin{equation}
	H(S,P) = \frac{1}{2} \sqrt{\frac{S}{\pi}} + \frac{4}{3} P S\sqrt{\frac{S}{\pi}} =  \sqrt{\frac{S}{\pi}} \bigg( \frac{1}{2} + \frac{4}{3} P S\bigg)~.
	\label{H}
\end{equation}
In this framework, the temperature is the state function
\begin{equation}
	T(S,P) = \frac{1}{4\sqrt{\pi S}} + 2P \sqrt{\frac{S}{\pi}}~,
\end{equation}
and the volume is
\begin{equation}
	V(S) = \frac{4\pi}{3} \left(\frac{S}{\pi}\right)^\frac{3}{2}~.
\end{equation}

Conversely, while the relation $dU = T dS - P dV$ is well-defined as a geometric identity, a traditional internal energy representation $U(S,V)$ is thermodynamically degenerate because the on-shell volume $V=V(S)$ depends solely on the entropy. Therefore, $S$ and $V$ cannot be treated as independent state variables. To circumvent this problem from a traditional thermodynamic perspective, we must work off-shell, meaning that we treat $S$, $P$, and $V$ as mutually independent variables. Later in this work, we will demonstrate how this expression can be properly understood from such an extended perspective.

The physical motivation for incorporating $P$ is clear, as non-zero pressure states describe different black hole solutions depending on the underlying matter content. In spherically symmetric vacuum spacetimes with no cosmological constant, the pressure vanishes, \mbox{$P = 0$}, and the formalism collapses to the classic Schwarzschild thermodynamics, yielding \mbox{$U = \sqrt{S/\pi}/2$} and \mbox{$T = 1/(4\sqrt{\pi S})$}. A well-known case with $P\neq 0$ is the Schwarzschild--(anti) de Sitter thermodynamics of black-hole chemistry, where the cosmological constant generates a pressure term \cite{mann2025black, dolan2011compressibility, dolan2011pressure, dolan2012pdv, caldarelli2000thermodynamics, gibbons2005first, cvetivc2011black, kubizvnak2012p, kubizvnak2017black, xiao2024extended, campos2024generating, kastor2009enthalpy}. A lesser-known but highly revealing example is the Reissner--Nordström black hole \cite{hayward1998unified}.

\subsection{Quasi-local thermodynamics of Reissner--Nordström}

Conventionally, the thermodynamics of the Reissner--Nordström solution is described by treating the electric charge~$Q$ as an independent variable, yielding the first law
\begin{equation}
	dm = T dS + \Phi dQ~,
\end{equation}
where $m$ is the mass parameter (which coincides with the ADM mass), $T$ is the Hawking temperature, and $\Phi$ is the electric potential evaluated at the horizon:
\begin{equation}
	m = \frac{r_+^2 + Q^2}{2r_+} ~, \qquad T = \frac{r_+^2 - Q^2}{4\pi r_+^3} ~, \qquad \Phi = \frac{Q}{r_+}~.
\end{equation}
However, this traditional picture does not coincide with the quasi-local formulation presented in the previous subsection. 

In the quasi-local framework, the internal energy representation is given by the Misner-Sharp mass evaluated at the horizon, yielding, in terms of the geometric parameters,
\begin{equation}
	U = m - \frac{Q^2}{2r_+}~.
\end{equation}
Moreover, the electromagnetic stress-energy tensor generates an effective negative pressure on the horizon,
\begin{equation}
	P = -\frac{Q^2}{8\pi r_+^4}~.
	\label{pressure RN}
\end{equation}
Conversely, in the enthalpy representation, the appropriate thermodynamic potential is given by the $H(S,P)$ of Eq.~\eqref{H}, or in terms of the geometric parameters,
\begin{equation}
	H = m - \frac{2 Q^2}{3r_+}~.
\end{equation}
It is worth noting that, within the enthalpy representation, the variation of the electric charge is entirely accounted for by the variation of the pressure~\eqref{pressure RN} in Eq.~\eqref{H_esf}.

While this unified quasi-local thermodynamic description is thoroughly established for spherically symmetric spacetimes, a generalized formulation incorporating both mechanical forms $dU = T dS - P dV$ and $dH = T dS + V dP$, while preserving a clear geometric volume, for rotating spacetimes, such as the Kerr--Newman black hole, has not, to our knowledge, been established. The geometric oblateness induced by rotation breaks the simple functional dependence between volume and entropy, requiring a fundamentally new approach. The purpose of this work is to bridge this gap. We present a quasi-local thermodynamic description for Kerr--Newman by extending the standard pressure and volume framework to include the shear deformation of the horizon.
The first step towards constructing this quasi-local thermodynamics for Kerr--Newman is to identify a suitable generalization of the Misner-Sharp mass to rotating geometries, which will serve as the internal energy. In the following section, we show that this can be rigorously achieved within the covariant phase space formalism.


\section{Internal energy via covariant phase space}
\label{sec:Iyer-Wald}

In the Iyer--Wald formalism, the variation of the Hamiltonian associated with a Killing field $\xi$ reduces to an integral over the boundary of a spacelike hypersurface $\Sigma$,
\begin{equation}
	\delta H_{\xi }\equiv\int_{\partial \Sigma }\left( \delta \mathbf{Q}\left[ \xi %
	\right] -\iota _{\xi }\boldsymbol{\Theta }\left( \phi ,\delta \phi \right)
	\right) ~,  \label{eqHbor}
\end{equation}
where $\phi$ represents the dynamical fields (including the metric), $\mathbf{Q}[\xi]$ is the Noether charge, and $\boldsymbol{\Theta}$ denotes the symplectic potential~\cite{wald1993black,iyer1994some}. To extract the Hamiltonian, its variation must be integrable. This requires the existence of a term $\boldsymbol{\mathcal{B}}(\phi)$ such that
\begin{equation}
	-\iota _{\xi }\boldsymbol{\Theta }\left( \phi ,\delta\phi \right) =	\delta\boldsymbol{\mathcal{B}}\left( \phi \right) ~. \label{Eqcomp}
\end{equation}

By restricting the formalism to the gravitational sector via the Einstein-Hilbert Lagrangian, the Noether charge takes the typical Komar form:
\begin{equation}
	Q_{\alpha\beta} = -\frac{1}{16\pi} \epsilon_{\alpha\beta\mu\nu} \nabla^\mu \xi^\nu~.
\end{equation}
Furthermore, within this restriction, it was recently shown~\cite{jaraba2026quasi} that, for generalized Kerr--Schild metrics $g_{\mu\nu} = \bar{g}_{\mu\nu} + f k_\mu k_\nu$ (where $\bar{g}_{\mu\nu}$ is an arbitrary fixed background, $f$ is a scalar, and $k^\mu$ is a null geodesic vector field), Eq.~\eqref{Eqcomp} is satisfied for
\begin{equation}
	\mathcal{B}_{\alpha \beta }=  -\frac{1}{16\pi }\epsilon _{\alpha \beta \mu \nu }\xi ^{\nu }\nabla _{\sigma }\bar{g}^{\sigma \mu } ~.
\end{equation}

By selecting $\xi^\mu$ as the stationary ($t^\mu$) and axisymmetric ($\varphi^\mu$) Killing vectors, one obtains the energy $\Delta_{\bar{g}}\mathcal{E}$ and angular momentum $\Delta_{\bar{g}}\mathcal{J}$ Hamiltonians relative to the fixed background $\bar{g}$. Integrating over a spatial 2-surface $\mathcal{S}$, these are given by
\begin{align}
	\Delta_{\bar{g}}\mathcal{E}& =-\frac{1}{8\pi}\int_{\mathcal{S}}\left(\nabla ^{\alpha}t^{\beta}+\nabla_{\sigma }\bar{g}^{\sigma \alpha}t^{\beta}\right) dS_{\alpha \beta}~,  \label{eq:DeltaE} 
	\\
	\Delta_{\bar{g}}\mathcal{J}& =\frac{1}{8\pi}\int_{\mathcal{S}}\left(\nabla^{\alpha}\varphi ^{\beta}+\nabla_{\sigma}\bar{g}^{\sigma\alpha}\varphi ^{\beta}\right) dS_{\alpha \beta}~,
	\label{eq:DeltaJ}
\end{align}
where $dS_{\alpha \beta }=n_{\alpha \beta }\,dudv$, with the binormal tensor $n_{\alpha \beta }$ defined in terms of the coordinates $(u,v)$ of $\mathcal{S}$ as 
\begin{equation}
	n_{\alpha \beta }\equiv \frac{1}{2}\epsilon _{\alpha \beta \mu \nu }\frac{%
		\partial x^{\mu }}{\partial u}\frac{\partial x^{\nu }}{\partial v}~.
\end{equation}

Let us consider a Kerr--Schild metric over a Minkowski spacetime ($\bar{g}_{\mu\nu} = \eta_{\mu\nu})$, which is conveniently written in oblate spheroidal coordinates $(t, r, \theta, \phi)$:
\begin{equation}
	d\bar{s}^{2} = -dt^{2} + \frac{\rho^{2}}{r^{2}+a^{2}}dr^{2} + \rho^{2} d\theta^{2} + (r^{2}+a^{2})\sin^{2}\theta d\phi^{2} ~, \label{Mink}
\end{equation}
with $\rho^{2} = \rho^2(r,\theta)=r^{2} + a^{2} \cos^{2}\theta$, where $a$ is the angular momentum parameter, and rotation is encoded in the 1-form
\begin{equation}
	k_{\mu}dx^{\mu} = dt + \frac{\rho^{2}}{r^{2}+a^{2}}dr - a \sin^{2}\theta d\phi ~.
\end{equation}

Several physically relevant stationary, axisymmetric black hole spacetimes are described by a Gürses--Gürsey metric, for which
\begin{equation}
	f(r,\theta )=\frac{2M\left( r\right) r}{\rho^2(r,\theta)}~.
	\label{eq:GG_profile}
\end{equation}
For this metric $g_{\mu\nu} = \eta_{\mu\nu} + f(r,\theta)\, k_\mu k_\nu$, considering $t^\mu = (\partial_t)^\mu$ and $\varphi^\mu = (\partial_\phi)^\mu$, it is found that~\cite{jaraba2026quasi}
\begin{align}
	\Delta_{\text{Mink}}\mathcal{E}_{\text{GG}} &= M(r) + \frac{ar - (r^{2}+a^{2}) \arctan(a/r)}{2a} M^{\prime}(r) ~, \label{eq:E_master} \\
	\Delta_{\text{Mink}}\mathcal{J}_{\text{GG}} &= aM(r) - \frac{(r^{2}+a^{2}) \left[ -ar + (r^{2}+a^{2}) \arctan(a/r) \right]}{2a^{2}} M^{\prime}(r) ~. \label{eq:J_master}
\end{align}
Eqs.~\eqref{eq:E_master} and~\eqref{eq:J_master} agree with the results obtained via pseudo-tensors in \cite{aguirregabiria2001energy}. However, as presented above, they also follow from the covariant approach of the Iyer--Wald formalism.

The position of the horizon $(r = r_+)$ is determined by
\begin{equation}
	f(r_+, \theta) = \frac{r_+^2+a^2}{\rho^2(r_+,\theta)}~,
\end{equation}
and it is generated by $\xi^\mu = (\partial_t)^\mu + \Omega_H (\partial_\phi)^\mu$, with the angular velocity
\begin{equation}
	\Omega_H = \frac{a}{r_+^2 + a^2}~.
\end{equation}
Evaluating the linear combination of the corresponding Hamiltonian charges~\eqref{eq:E_master} and~\eqref{eq:J_master} along this generator on the horizon, we find the remarkably simple result 
\begin{equation}
	U \equiv \Delta _{\text{Mink}}\mathcal{E}_{\text{GG}} - \Omega_H \Delta _{\text{Mink}}\mathcal{J}_{\text{GG}} = \frac{r_+}{2}~.
\end{equation}

The emergence of this result for such a broad variety of Kerr--Schild spacetimes supports identifying $U = r_+/2$ as a generalized Misner-Sharp mass for rotating geometries.
Extending the spherically symmetric case, we interpret this generalized Misner-Sharp mass as the thermodynamic internal energy of the black hole. This requires assigning the homogeneity weights~\eqref{scaling}, with the corresponding Euler relation,
\begin{equation}
	U = 2TS - 3PV~.
	\label{euler}
\end{equation}

Within the class of spacetimes considered, the pressure-volume term
in Eqs.~\eqref{eq:DeltaE} and~\eqref{eq:DeltaJ} comes from
\begin{equation}
	\left(\nabla_\alpha\eta^{\alpha[\beta}\right)\xi^{\gamma]}-\nabla^\beta \xi^\gamma
	=8\pi r\left(T^r{}_r-\rho_\Lambda\right)\varepsilon^{\beta\gamma}+\mathcal{O}(a)~,
	\label{PVKS}
\end{equation}
where $\varepsilon^{\beta\gamma}=t^\beta k^\gamma-k^\beta t^\gamma$, and $\mathcal{O}(a)$ denotes a tensor that vanishes as $a\to0$. In spherical symmetry, we identify the pressure defined in Eq.~\eqref{PT}. For rotating configurations, the remainder describes a rotational contribution.

It may be the case that the standard differential relation $dU = T dS - P dV$ is not satisfied for a particular black hole, that is, a choice of scalar profile~\eqref{eq:GG_profile}. Then, the first law must be extended by introducing additional work terms associated with the geometric parameters of the metric, such as $r_+$, the rotation parameter $a$, and the electric charge $Q$. Thus, in an appropriately extended configuration space, one can introduce additional independent variables $Y_i=Y_i(r_+,a,Q,\dots)$ and their thermodynamic conjugates $X_i$ such that
\begin{equation}
	dU = T dS - P dV + \sum_i X_i dY_i~.
\end{equation}

Although this is algebraically possible, for this extended first law to remain compatible with the Euler relation, the additional variables $Y_i$ must have a vanishing scaling weight, ensuring that their corresponding work terms do not contribute to the homogeneity scaling.

Importantly, there is no general guarantee that the discrepancy between $dU$ and $T dS-PdV$ can be absorbed solely by differentials of dimensionless variables. Thus, finding a set of dimensionless $Y_i$ that renders the first law exact while preserving the Euler relation is a non-trivial requirement. Moreover, even when such variables exist, they must possess a clear physical or geometric interpretation.

In the following section, we show that a single dimensionless parameter $Y$ is sufficient to achieve this for Kerr--Newman geometries. This addition matches the number of thermodynamic degrees of freedom to the three independent geometric parameters of the spacetime $(r_+, a, Q)$. However, this choice of $Y$ is not unique, since any reparameterization $Y\to\widetilde{Y}(Y)$ leaves the work term invariant provided its thermodynamic conjugate is redefined as $\widetilde{X} = X dY/d\widetilde{Y}$. Hence, we identify a case with clear geometric interpretation, for which both the extended first law and the Euler relation are satisfied.


\section{Quasi-local thermodynamics of Kerr--Newman}
\label{sec:quasi-local Kerr--Newman}

\subsection{Internal energy and enthalpy representations}
\label{subsec:quasi-local KN U}

The Kerr--Newman black hole represents an important application of the geometric structure and thermodynamic interpretation developed in the previous section. It is characterized by the mass profile \mbox{$M(r)=m-Q^{2}/(2r)$} in Eq.~\eqref{eq:GG_profile}. For this spacetime, the Euler relation~\eqref{euler} holds with the following Hawking temperature, entropy, and geometric volume \cite{poisson2004relativist,ballik2013vector}:
\begin{equation}
	\begin{split}
		T &= \frac{r_+^2-a^2-Q^2}{4\pi r_+(r_+^2+a^2)}~,\qquad
		S = \pi(r_+^2+a^2)~,\\[4pt]
		V &= \frac{4\pi}{3}\,r_+(r_+^2+a^2)~.
	\end{split}
	\label{eq:TSV_KN}
\end{equation}
From Eq.~\eqref{euler}, the quasi-local pressure is fixed as
\begin{equation}
	P = -\frac{a^2+Q^2}{8\pi r_+^2(r_+^2+a^2)}~.
	\label{eq:P}
\end{equation}
A pressure term depending on the angular momentum parameter is also derived in the surface-tension formalism \cite{hansen2016criticality, chen2017modified, chen2019modifying, villalba2020kerr, jawad2018thermodynamic, fernandez2026penrose}. However, it does not coincide with~\eqref{eq:P}, since in that approach the thermodynamic volume does not have a clear geometric interpretation. This is in contrast to $V$ in~\eqref{eq:TSV_KN}, which is the volume within an ellipsoid of revolution \cite{ballik2013vector}. In contrast to the spherically symmetric case, Eq.~\eqref{eq:P} reveals a nonzero pressure even in vacuum (Kerr, $Q=0$) due to the rotational correction $\mathcal{O}(a)$ in Eq.~\eqref{PVKS}.

A direct computation, performed by expanding the differentials in terms of the geometric parameters $(r_+, a, Q)$, verifies that the first law holds as
\begin{equation}
	dU = TdS - PdV + XdY~,
	\label{eq:1st_law_exact}
\end{equation}
where $Y$ is the (second) eccentricity parameter and $X$ is its
thermodynamic conjugate,
\begin{equation}
	Y = \frac{a}{r_+}~,\qquad
	X = -\frac{a\left(3r_+^2-a^2-Q^2\right)}{6\left(r_+^2+a^2\right)}~.
	\label{eq:XY}
\end{equation}

To show that $Y$ is indeed an eccentricity parameter, it is sufficient to evaluate the horizon surface $r = r_+$ in the Cartesian coordinates of the Minkowski background \eqref{Mink}, which yields
\begin{equation}
	\frac{x^2 + y^2}{r_+^2 + a^2} + \frac{z^2}{r^2_+} = 1~.
\end{equation}
The first and second eccentricities of this ellipsoid are \mbox{$\widetilde Y = a/\sqrt{r_+^2 + a^2}$} and $Y = a/r_+$, respectively.

Since $Y$ is a ratio of lengths, it is scale-invariant under Euler
transformations (weight zero). Thus, the product $XY$ does not
contribute to the Smarr relation~\eqref{euler}, preserving the original
homogeneity. Hence, accommodating the loss of spherical symmetry amounts to
the first law acquiring an additional work term associated with the
geometric deformation of the horizon.

To establish Eq.~\eqref{eq:1st_law_exact}, the system must be formulated off-shell, treating $S$, $V$, and $Y$ as mutually independent variables, with $V$ coinciding with the geometric volume $V(S,Y)$ of~\eqref{eq:TSV_KN} only on-shell. To construct this extension unambiguously, we first work in the enthalpy representation,
\mbox{$H = U + PV$}:
\begin{equation}
	H = \frac{r_+}{2} - \frac{a^2+Q^2}{6r_+}~.
	\label{eq:H_geom}
\end{equation}
Since $(S,Y,P)$, regarded as functions of $(r_+,a,Q^2)$, are smooth and define a bijective map, satisfying
\begin{equation}
	\frac{\partial(S,Y,P)}{\partial(r_+,a,Q^2)} = -\frac{1}{4r_+^4} \neq 0~,
	\label{eq:jacobian}
\end{equation}
they constitute a genuine coordinate system. Consequently, $H$ is a
well-defined, single-valued function of the independent variables $(S,P,Y)$:
\begin{equation}
	H(S,P,Y) = \sqrt{\frac{S}{\pi(1+Y^2)}}\left(\frac12+\frac43 PS\right)~,
	\label{eq:H_SPY}
\end{equation}
with first law
\begin{equation}
	dH = TdS + VdP + XdY~.
	\label{eq:1st_law_H}
\end{equation}
The equations of state follow by direct differentiation with respect to the
independent variables $(S,P,Y)$:
\begin{align}
	T(S,P,Y) &= \left(\frac{\partial H}{\partial S}\right)_{P,Y}
	= \frac{1}{4\sqrt{\pi S(1+Y^2)}} + 2P\sqrt{\frac{S}{\pi(1+Y^2)}}~, \label{THS}
	\\
	V(S,Y) &= \left(\frac{\partial H}{\partial P}\right)_{S,Y}
	= \frac{4S}{3}\sqrt{\frac{S}{\pi(1+Y^2)}}~,
	\label{VHP}
	\\
	X(S,P,Y) &= \left(\frac{\partial H}{\partial Y}\right)_{S,P}
	= -\frac{Y}{1+Y^2}\sqrt{\frac{S}{\pi(1+Y^2)}}\left(\frac12+\frac43 PS\right)~.
	\label{XHY}
\end{align}
As in the spherically symmetric case, the variation of the electric charge is completely absorbed into $dP$. 

While the enthalpy representation is direct, the internal-energy
representation requires care. The origin of the subtlety is that $H$ is
linear in $P$,
\begin{align}
	H(S,P,Y) &= U(S,Y) + V(S,Y)P~, 
	\label{eq:H_linear}
	\\
	U(S,Y) &= \frac12\sqrt{\frac{S}{\pi(1+Y^2)}}~,
\end{align}
so that $\partial H/\partial P$, Eq.~\eqref{VHP}, is independent of $P$. Consequently, the map $(S,P,Y)\mapsto(S,V,Y)$ is degenerate, its Jacobian
vanishes, and a function $P=P(S,V,Y)$ cannot be constructed. 

We circumvent this by defining $U = H - PV$ prior to imposing the on-shell condition
\mbox{$V=\partial H/\partial P = V(S,Y)$}. Using Eq.~\eqref{eq:H_linear},
\begin{equation}
	U= U(S,Y) + P\left[V(S,Y) - V\right]~,
	\label{eq:U_offshell}
\end{equation}
which is well-defined on the off-shell, extended, four-dimensional configuration space
$(S,P,V,Y)$ with these variables mutually independent. The differential of Eq.~\eqref{eq:U_offshell}
is
\begin{equation}
	dU = TdS + \big[V(S,Y)-V\big]dP - PdV + XdY~,
	\label{eq:dU_offshell}
\end{equation}
with
\begin{equation}
	\left(\frac{\partial U}{\partial S} \right)_{P,V,Y} = \left( \frac{\partial H}{\partial S} \right)_{P,Y}~, \qquad \left( \frac{\partial U}{\partial Y}\right)_{S,P,V} = \left( \frac{\partial H}{\partial Y} \right)_{S,P}~.
\end{equation}
Only upon imposing the on-shell condition
$V=V(S,Y)$ does the middle term vanish, and
Eq.~\eqref{eq:dU_offshell} reduces to Eq.~\eqref{eq:1st_law_exact}. In Eq.~\eqref{eq:U_offshell}, $U$ is not an
internal energy in the traditional sense, but rather the unique function satisfying ($i$) $\partial U/\partial V = -P$, and ($ii$) reducing on-shell to $U(S,Y)$. In short, condition ($i$) implies $U = -PV + C(S,P,Y)$, while ($ii$) fixes $C = U(S,Y) + P V(S,Y)$~.

\subsection{The Reissner--Nordström and Kerr limits}
In the present subsection, we analyze the relevant limit configurations. The Reissner--Nordström case ($a=0$) is straightforward and reduces identically to the spherically symmetric framework discussed in Section~\ref{sec:quasi-local spherical}.

For pure Kerr ($Q=0$), the standard thermodynamic parameters are \cite{bardeen1973four,poisson2004relativist}
\begin{align}
	\begin{split}
		m &= \frac{r_+^2 + a^2}{2r_+}~, \qquad J = m a = \frac{a(r_+^2 + a^2)}{2r_+}~, \\
		\Omega &= \frac{a}{r_+^2 + a^2}~.
	\end{split}
\end{align}
The Hawking temperature and entropy coincide with Eq.~\eqref{eq:TSV_KN} in the appropriate limit. With these choices, the classical first law of thermodynamics for the Kerr black hole is $dm = T dS + \Omega dJ$.

The product $\Omega J$ represents the energy associated with rotation. Subtracting this rotational contribution from the total mass, we exactly recover the internal energy:
\begin{equation}
	m - \Omega J = \frac{r_+^2 + a^2}{2r_+} - \frac{a^2}{2r_+} = \frac{r_+}{2} = U~.
	\label{eq:M_omega_J}
\end{equation}

It is a straightforward computation that
\begin{equation}
	- P dV + X dY = - J d\Omega~.
\end{equation}
Therefore, in our model, the mechanical work of spatial volume deformation combined with the variation of the eccentricity parameter $Y$ together encode the rotational work $-J d\Omega$, which is characteristic of an ensemble where the angular velocity acts as a thermodynamic variable. 

In other words, for the pure Kerr limit, the internal energy~$U$ can be understood simply as a Legendre transformation of the mass parameter~$m$ with respect to the rotational variables, followed by a reparameterization. This correspondence holds because pure Kerr is a vacuum solution, and, thus, the global asymptotic charges $m$ and $J$ coincide with the quasi-local charges evaluated at the horizon. On the other hand, for the Kerr--Newman spacetime, the quasi-local formulation cannot be interpreted analogously. Due to the presence of the electromagnetic field, the quasi-local charges~\eqref{eq:E_master} and~\eqref{eq:J_master} are functions of the radial position of the 2-surface of integration. Consequently, the internal energy~$U$ cannot be identified with a simple Legendre transform of the global asymptotic charges.

\subsection{Shear deformation of the horizon}

In the preceding subsections, the dimensionless parameter $Y$ was introduced to account for the rotational deformation of the black hole, successfully absorbing part of the rotational work into the first law. To rigorously ground this thermodynamic extension geometrically, we must analyze the kinematic deformation of the two-dimensional induced metric on the horizon surface.

The components of this transverse metric can be written exactly in terms of our independent variables in the form
\begin{align}
	q_{\theta\theta}(S, Y, \theta) &= S \left[ \frac{1 + Y^2 \cos^2\theta}{\pi(1+Y^2)} \right], \\
	q_{\phi\phi}(S, Y, \theta) &= S \left[ \frac{1+Y^2}{\pi(1+Y^2 \cos^2\theta)} \sin^2\theta \right]~.
\end{align}
Notably, the metric takes the form $q_{AB}(S, Y,\theta) = S \, \tilde{q}_{AB}(Y, \theta)$, where the entropy acts as a conformal factor.

Let us consider the total variation of the horizon metric induced by a thermodynamic perturbation:
\begin{equation}
	\delta q_{AB} = \left( \frac{\partial q_{AB}}{\partial S} \right)_Y \delta S + \left( \frac{\partial q_{AB}}{\partial Y} \right)_S \delta Y~.
\end{equation}
Due to the factorization highlighted above, the partial derivative with respect to $S$ is immediate, 
\begin{equation}
	\left( \frac{\partial q_{AB}}{\partial S} \right)_Y \delta S = \tilde{q}_{AB} \, \delta S = \left( \frac{\delta S}{S} \right) q_{AB}~.
\end{equation}
Being strictly proportional to the background metric $q_{AB}$, this term configures a pure isotropic expansion of the two-dimensional metric \cite{poisson2004relativist}.

To classify the remaining geometric deformation governed by $Y$, we calculate its trace using Jacobi's formula,
\begin{equation}
	q^{AB} \left( \frac{\partial q_{AB}}{\partial Y} \right)_S \delta Y = \frac{\partial \ln q}{\partial Y} \delta Y~,
\end{equation}
where $q \equiv \det(q_{AB})$. Calculating the determinant of the metric in the variables $(S, Y)$,
\begin{equation}
	q = q_{\theta\theta} q_{\phi\phi} = \left( \frac{S^2}{\pi^2} \right) \sin^2\theta~.
\end{equation}
The determinant $q$ depends exclusively on the entropy $S$ and the angular coordinate $\theta$, being independent of the shape parameter~$Y$. Consequently,
\begin{equation}
	\frac{\partial \ln q}{\partial Y} = 0 ~,
\end{equation}
and thus
\begin{equation}
	q^{AB} \left( \frac{\partial q_{AB}}{\partial Y} \right)_S \delta Y = 0~.
\end{equation}
This proves that the perturbation associated with $\delta Y$ generates a symmetric traceless tensor, which, in surface kinematics, characterizes a shear deformation \cite{poisson2004relativist}.

Therefore, the variation $\delta Y$ governs the oblate distortion of the black hole. It effectively flattens the poles and stretches the equator of the horizon while maintaining both the entropy $S$ and, therefore, the surface area strictly constant. In the framework of continuum mechanics, a work term associated with this type of deformation is driven by a shear force. Consequently, the conjugate variable $X$ can be interpreted as a thermodynamic shear tension on the horizon.

\section{Final Remarks}
\label{sec:final_remarks}

In this work, we established a quasi-local thermodynamic framework for the Kerr--Newman black hole. Our formalism was built upon the general quasi-local thermodynamics of spherically symmetric horizons based on pressure and volume terms. We demonstrated that accommodating the geometric deformation of the horizon caused by rotation requires extending the configuration space with an eccentricity parameter $Y$. This extension manifests as a shear work term $XdY$ in the first law, where $X$ is a thermodynamic shear tension.

The enthalpy representation offers a more natural and direct approach, where the thermodynamic variables are independent. Conversely, working within the internal energy representation reveals the necessity of treating the volume $V$ as an auxiliary independent variable in an off-shell formulation. This distinguishes it from the geometric volume, which is constrained by the horizon geometry as a state function $V(S,Y)$. 

Within this treatment, the extended phase space achieves a crucial physical decoupling. It separates the traditional mechanical work $-P dV$, associated with the internal thermodynamic pressure, from the purely two-dimensional dynamics of the horizon surface: the isotropic area expansion $T dS$ (the thermal energy) and the area-preserving shear deformation $X dY$. 

In summary, our framework generalizes the quasi-local thermodynamic description of spherically symmetric horizons to include the effects of rotation. 
This approach may offer insights into new formulations of the Iyer--Wald formalism \cite{wald1993black, iyer1994some, campos2025black, xiao2024extended} and the quantum statistical relation \cite{gibbons1977action, gibbons1978black}. Moreover, it broadens the scope of black-hole chemistry~\cite{mann2025black, kubizvnak2017black} by utilizing variables that are fundamentally rooted in classical thermodynamics. 
The surface-tension approach~\cite{hansen2016criticality} has been applied to a variety of other horizons~\cite{chen2017modified, chen2019modifying, villalba2020kerr, jawad2018thermodynamic}, and has recently been connected to the Newman-Penrose and Geroch-Held-Penrose formalisms~\cite{fernandez2026penrose}. Extending our framework to these cases is also a natural direction for future work.


\section*{Acknowledgements}
T.~L.~C. acknowledges the support of Funda\c{c}\~ao Carlos Chagas Filho de Amparo \`{a} Pesquisa do Estado do Rio de Janeiro (FAPERJ) -- Brazil, Grant No.~E-26/202.775/2026 and 202.776/2026.










\end{document}